\documentclass{IEEEtran}
\pdfoutput=1
\usepackage{amsmath}
\usepackage{amssymb}
\usepackage{cite}
\usepackage{graphicx}
\usepackage{subfig}
\usepackage[nolist]{acronym}
\usepackage{booktabs}
\usepackage{comment}
\usepackage{flushend}
\usepackage{paralist}
\usepackage{color}
\usepackage{soul}
\usepackage{tikz}
\usepackage{pgfplots}
\usepackage{bm}
\pgfplotsset{compat=newest}
\usepackage{slashbox}
\usepackage[normalem]{ulem}

\definecolor{orange}{cmyk}{0,0.5,1,0}

\iffalse
\newcommand{\IGa}[1]{{\color{blue}#1}} 
\else
\newcommand{\IGa}[1]{#1} 
\fi

\newcommand{\figref}[1]{Figure~\ref{#1}}

\newcommand{\tabref}[1]{Table~\ref{#1}}

\newcommand{\specialcell}[2][c]{%
	\begin{tabular}[#1]{@{}c@{}}#2\end{tabular}}

\begin{document}
\title{GFDM - A Framework for Virtual PHY Services in 5G Networks}

\author{
	\IEEEauthorblockN{
		\normalsize{
            Ivan~Gaspar\IEEEauthorrefmark{1},
			Luciano~Mendes\IEEEauthorrefmark{2},
			Maximilian~Matth\'e\IEEEauthorrefmark{1},
            Nicola~Michailow\IEEEauthorrefmark{1},
            Dan~Zhang\IEEEauthorrefmark{1},
			Antonio~Alberti\IEEEauthorrefmark{2}
            Gerhard~Fettweis\IEEEauthorrefmark{1}
		}
	}\\
	\IEEEauthorblockA{
		\small{
			\IEEEauthorrefmark{1}Vodafone Chair Mobile Communication Systems, Technische Universit\"at Dresden, Germany\\
			\IEEEauthorrefmark{2}Instituto Nacional de Telecomunica\c{c}\~{o}es, Sta. Rita do Sapuca\'{\i}, MG, Brazil\\
			\texttt{\{first name.last name\}@ifn.et.tu-dresden.de,
                        \{luciano|alberti\}@inatel.br}}
		}
	}
\maketitle


\begin{acronym}
  \acro{2G}{2nd generation}
  \acro{3G}{3rd generation}
  \acro{4G}{4th generation}
  \acro{5G}{5th generation}
  \acro{APP}{a posteriori probability}
  \acro{AS}{Access Stratum}
  \acro{ASIP}{Application Specific Integrated Processors}
  \acro{AWGN}{additive white Gaussian noise}
  \acro{BS}{base stations}
  \acro{CP}{cyclic prefix}
  \acro{CR}{Cognitive Radio}
  \acro{CSMA}{carrier sense multiple access}
  \acro{CoMP} {Coordinated Multipoint}
  \acro{COQAM}{cyclic OQAM}
  \acro{CB-FMT}{cyclic block filtered multitone}
  \acro{DFT}{discrete Fourier transform}
  \acro{DL}{downlink}
  \acro{EPC}{evolved packet core}
  \acro{FBMC}{Filterbank multicarrier}%
  \acro{FDMA}{frequency division multiple access}
  \acro{FPGA}{field programmable gate array}
  \acro{FTN}{Faster than Nyquist}%
  \acro{FT}{Fourier transform}
  \acro{FMT}{filtered multitone}
  \acro{GFDM}{Generalized Frequency Division Multiplexing}
  \acro{ICI}{intercarrier interference}
  \acro{IDMA}{interleave division  multiple access}
  \acro{IMS}{IP multimedia subsystem}
  \acro{I}{in-phase}
  \acro{IoT}{Internet of Things}
  \acro{IP}{Internet Protocol}
  \acro{ISI}{intersymbol interference}
  \acro{IUI}{inter-user interference}
  \acro{IOTA}{isotropic orthogonal transform algorithm}
  \acro{LUT}{look-up table}
  \acro{LTE}{Long Term Evolution}
  \acro{M2M}{Machine-to-Machine}
  \acro{MA}{multiple access}
  \acro{MC}{multicarrier}
  \acro{MIMO}{multiple-input multiple-output}
  \acro{MS}{mobile stations}
  \acro{MSE}{mean-squared error}
  \acro{MTC}{machine type communication}
  \acro{NFV}{network functions virtualization}
  \acro{NOMA}{non-orthogonal multiple access}
  \acro{OFDM}{Orthogonal Frequency Division Multiplexing}
  \acro{OOB}{out-of-band}
  \acro{OSS}{Operations Support Systems}
  \acro{OQAM}{offset quadrature amplitude modulation}
  \acro{PAPR}{peak-to-average power ratio}
  \acro{PHY}{physical layer}
  \acro{Q}{quadrature-phase}
  \acro{RC}{raised cosine}
  \acro{RF}{radio frequency}
  \acro{RRC}{root raised cosine}
  \acro{SDNs}{software-defined networks}
  \acro{SDN}{software-defined networking}
  \acro{SDR}{software-defined radio}
  \acro{SDW}{software-defined waveform}
  \acro{QAM}{quadrature amplitude modulation}
  \acro{B-OFDM}{block OFDM}
  \acro{SC-FDE}{single carrier with frequency domain equalization}
  \acro{SC-FDM}{single carrier frequency division multiplexing}
  \acro{SCMA}{sparse code multiple access}
  \acro{SEFDM}{spectrally efficient frequency division multiplexing}
  \acro{SER}{symbol error rate}
  \acro{STFT}{short-time Fourier transform}
  \acro{TR-STC}{time-reversal space-time coding}
  \acro{UFMC}{Universal Filtered Multicarrier}
  \acro{UL}{uplink}
  \acro{V-OFDM}{Vector OFDM}
  \acro{ZF}{zero-forcing}
  \acro{ZP}{zero-padding}
  \acro{WLAN}{wireless Local Area Network}
  \acro{WRAN}{Wireless Regional Area Network}
  \acro{XaaS}{anything or everything as a service}
  \acro{QoS}{quality of service}
\end{acronym}

\begin{abstract}
The next generation of wireless networks will face different challenges from new scenarios. The main contribution of
this paper is to show that \IGa{\ac{GFDM}, as a baseline of flexible circular filtered multicarrier systems, can be used as a framework to virtualize the \acs{PHY} service} for the upper layers of 5G networks. This framework opens the possibility to apply software-defined network
principles to produce software-defined waveforms capable of addressing the requirements of future mobile networks. Hence, \IGa{a block oriented}
concept will be used to provide the modulation service, emulating \IGa{different flavors of waveforms designed to go beyond the well-established \ac{OFDM} principles,} in scenarios where they perform best.
The virtual \ac{PHY} service opens the opportunity to have a fast and dynamic evolution of the infrastructure, as
applications change over time. \IGa{The presented unified modulation concept contributes with future research directions to} address burst and continuous transmissions\IGa{, referencing basic approaches for synchronization and advanced receiver design that can be exploited in future for the whole frame structure design and channel estimation strategies.}
\end{abstract}

\section{Introduction}\label{sec:introduction}

\begin{figure*}[t]
  \centering
  \includegraphics[width=14.5cm]{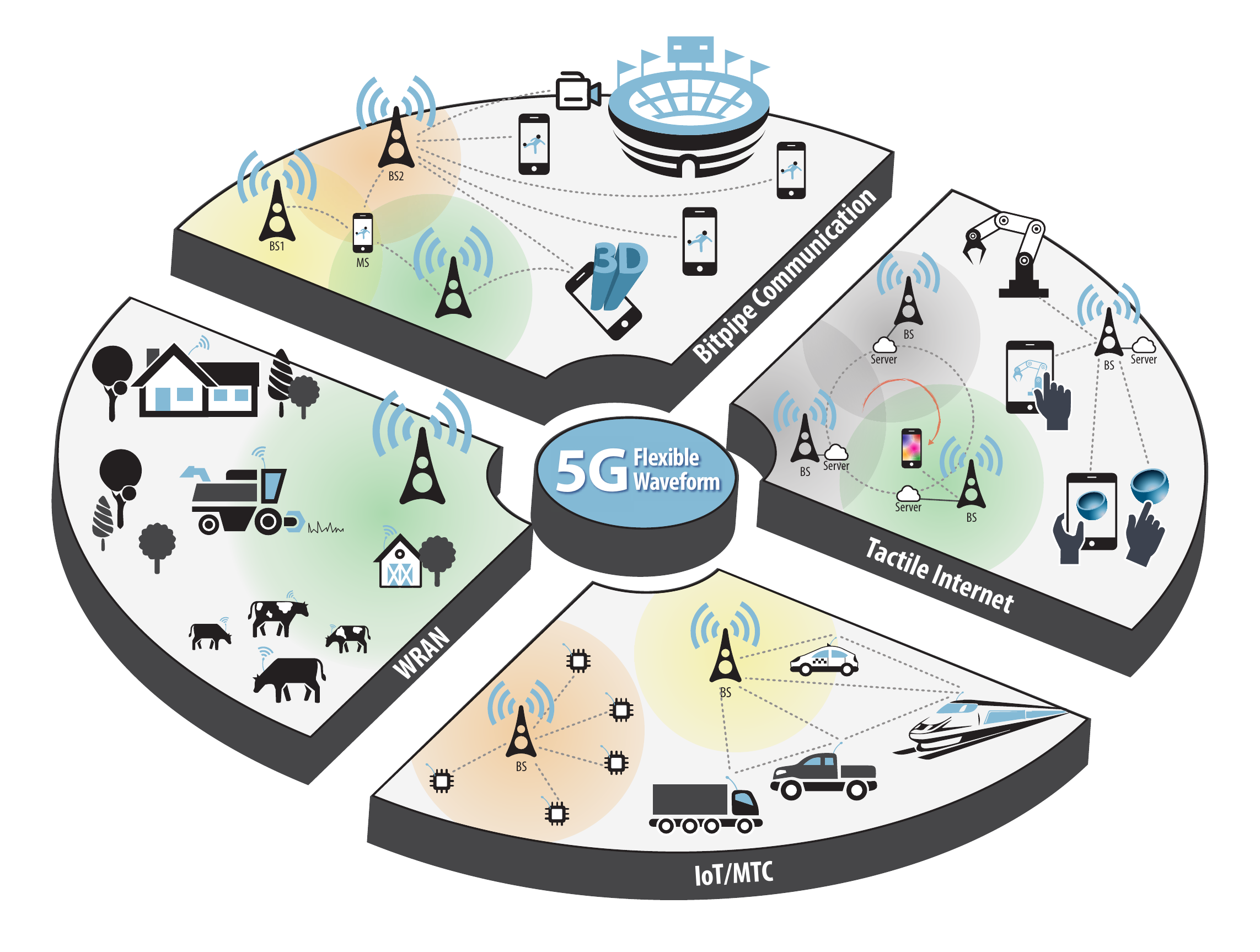}
  \caption{Main scenarios for 5G networks.}
  \label{fig:scenarios}
\end{figure*}

\IEEEPARstart{T}{he} role of software in mobile communication systems has increased with time. The \ac{2G} incorporated  several \ac{OSS} implemented at computer servers, allowing operators to better manage their networks. In the \ac{3G}, dozens of functionalities have been implemented on computing systems, particularly those employed to convey voice over the \ac{IP}. The \ac{4G} continued taking advantage of the increased capacity of computing systems and their reduced cost to implement more functions at the software level. For the upcoming \ac{5G}, the application of \ac{SDN} to mobile networks can ease the network management by enabling \emph{anything as a service}.

Today, \ac{BS} and, in some cases, even \ac{MS} are becoming software-defined and even virtualized. For instance, \ac{PHY} functionalities are being defined by local control software. Additionally, \ac{SDR} enables  radio virtualization, where several radio components are implemented in software. \ac{CR} goes one step further with a software-based decision cycle to self-adapt \ac{SDR} parameters and consequently optimize the use of the communication resources. This proposal triggered the possibility of having real-time communication functionalities at virtual machines in cloud computing data centers, instead of deploying specialized hardware. The \ac{NFV} claims for cloud-based virtualization of network functionalities. The perspective is that all these software-based concepts should converge while designing \ac{5G} networks. A new breakthrough will be achieved when all these software paradigms are applied to the \ac{PHY} layer.

The 5G \ac{PHY} requires unprecedented levels of flexibility, performance, reliability, efficiency, robustness, energy-awareness, and scalability. When considering the scenarios for \ac{5G}, namely: \emph{\ac{IoT}}, \emph{Tactile Internet}, \emph{bitpipe connectivity}, and \emph{\ac{WRAN}}, the requirements go far beyond increasing throughput: loose synchronization for \ac{IoT}; low latency for Tactile Internet; reliable, efficient, and robust high throughput for bitpitpe communication; high coverage and dynamic spectrum allocation with low \ac{OOB} emission and \ac{CR} techniques to cover \ac{WRAN} applications. Certainly, all these scenarios will benefit from \ac{MIMO}-techniques, such as increased data rate, enhanced robustness and the additional degree of freedom in space for multi-user und multi-cell interference management. In order to maximize MIMO profits, the 5G \ac{PHY} must be able to cope with multi-antenna interference and related algorithms need to be implemented with an affordable complexity. 

5G related topics have been intensely researched lately and several projects were funded by the European Union \cite{Pirinen2014}, e.g. METIS, 5GNOW and now fantastic5G. It is widely believed that the fundamental characteristics of the main \ac{MC} system used today, i.e. \ac{OFDM}, are no longer compliant to emergent \ac{QoS} requirements. For instance, its high \ac{OOB} emission prevents its use in \ac{CR} and \ac{WRAN}. Therefore, \emph{waveform design} has been a key 5G research topic\IGa{, specially for the frequency spectrum below 6~GHz, which has already been reserved to a large extent for various legacy systems \cite{Pirinen2014}}. Given its low \ac{OOB} emissions, \ac{FBMC} \cite{Banelli2014} was rediscovered for \ac{CR} and dynamic spectrum allocation. On the other hand, the long impulse response of the filters, typically leading to the overlapping of at least 4 data symbols, prohibits its use for applications with sporadic traffics and tight latency constraints. Another waveform candidate is \ac{FTN} signaling \cite{Banelli2014}. Taking advantage of the Mazo limit, it is a promising solution for high data rate scenarios. But the large complexity of the receiver makes it unsuitable for \ac{IoT}.

While tailoring one specific waveform for each \ac{5G} scenario, it is much more desirable to adopt a single flexible waveform that can be easily reconfigured to address a multitude of applications. More importantly, such a \ac{SDW} builds a foundation at \ac{PHY} for preparing the aforementioned paradigm shift towards software-defined virtualization. Namely, the \ac{SDW} shall be generated in programmable hardware, based on \ac{ASIP}, \ac{FPGA}, or software with manageable cost. By means of a cost effective approach that exposes the time-frequency resource grid and waveform engineering capabilities to software, it becomes feasible to customize \ac{PHY} such that it can be seen as a virtual service for upper layers.

In this context,  the \ac{GFDM}  \IGa{pioneer concept of a circular filtered \ac{MC} system \cite{Michailow2014}, along with other waveforms presented in Section \ref{sec:waveforms},
}
provides a very flexible time-frequency structure that favors software exposition, controlling, and virtualization. \IGa{These circular filtered \ac{MC} system are} based on a number of independent modulated blocks formed by subcarriers and subsymbols. A software exposed transceiver \IGa{using this principle} enables local or logically centralized control of communication resources. This allows a software node to determine the best communication solution for a certain scenario. Alternatively, as in current \ac{SDN}, a logically centralized software controller can define transceiver configurations to address the demands of several different users. In this sense, the main goal of this paper is to explore the \IGa{flexibiblity of filtered \ac{MC} systems} to achieve a framework that covers all major waveform candidates for the different \ac{5G} scenarios, leading to software-defined, virtualized \ac{5G} networks. We will show that the main waveforms considered for \ac{5G} are, in fact, corner cases \IGa{of each other}. Also, the flexibility offered by \IGa{circular filtered \ac{MC} systems} is sufficient and necessary to fulfill diverse \ac{QoS} requirements in 5G. 

The remaining sections are organized as follows: Section~II describes the 5G scenarios. Section III links the \IGa{waveform design} to Gabor analysis. Section~IV introduces main aspects of GFDM, while Section~V shows how offset modulation can enhance the flexibility of the waveform. Section VI presents how \IGa{GFDM, as baseline,} can be \IGa{used} to \IGa{describe the different flavors that can be achieved with the circular filtered systems to support the \ac{5G} \ac{PHY}}. Section VII enlightens how a unified service framework can benefit from GFDM flexibility to provide multiple services. \IGa{Section VIII briefly discusses new research issues with respect to frame design and presents directions to design a corresponding flexible receiver}. Finally, Section IX concludes the paper.

\section{5G Scenarios: What to expect next?}\label{sec:scenarios}

A major difference between \ac{5G} and previous generations of mobile networks is the diversity of requirements that must be addressed for different applications, which are presented in \figref{fig:scenarios}. New challenges are on the horizon and for the first time increasing throughput is not the final answer. Of course, dense content (such as 3D or 4k videos) will demand even higher data rates. Hence, \emph{bitpipe communication} will still play an important role in the next generation of mobile communication networks. In this scenario, a \ac{PHY} with high spectrum and energy efficiency should be used by small sized, densely deployed cells. \ac{CoMP} \cite{Wunder2014}, which is a set of algorithms that uses information from all overlapping \ac{BS} to mitigate multicell interference, will be mandatory to achieve the required throughput for bitpipe communication.

One completely new aspect for 5G networks is the very low latency. Recent \ac{4G} deployment has been optimized for latency around 20 ms. However, new concepts as \emph{Tactile Internet} \cite{Fettweis2014} demand latencies that are, at least, one order of magnitude below this target. Tactile Internet is redefining `fast mobile Internet', where download data rates of hundreds of Mbps do not solve the problem. While today all major applications are hosted and running in the cloud, users want to maintain the responsiveness of locally executed software. Hence, besides having a low latency communication chain, the cloud servers must be close to the users and even follow the users while they move from one cell to another. New wearable hardware are also pushing for a latency reduction of 5G networks. Clearly, the \ac{PHY} is not the only bottleneck for latency in the communication chain. Huge efforts must be put into the \ac{AS}, responsible for transporting data over the wireless connection and managing radio resources, and into the development of the upper layers as well. Besides a short \ac{PHY} frame, Fog computing \cite{bonomi2014fog} must be exploited in order to achieve a low latency solution for Tactile Internet.

\emph{\ac{IoT}}, based on \emph{\ac{MTC}}, is considered as the next killer application for mobile networks. Although a business model has not yet raised, measurements and monitoring for health, transportation, home appliances, smart grid, etc. are pushing the development of solutions for \ac{MTC} and \ac{IoT}. From the \ac{PHY} point-of-view, the biggest challenge in this scenario is to provide coverage for a multitude of devices that have limited energy and processing capacity. The \ac{5G} waveform must allow these devices to send and receive data without being fully synchronized with the \ac{BS}, so that minimal energy is spent to achieve synchronization. The \ac{5G} waveform  must have very low \ac{OOB} emission, must be robust against time misalignment and must allow reliable one shot communication.

One final frontier for wireless communication is the coverage of low populated areas using \emph{\ac{WRAN}}. Today's technologies cannot offer proper Internet access for those who live in remote areas. While wired technologies have limited coverage, mobile communication networks are limited to a relatively small cell size and require licensed frequencies, rendering them economically unfeasible for this purpose. IEEE 802.22 has been proposed to address this scenario by opportunistically using vacant TV channels. \IGa{In addition to IEEE 802.22, there is also an effort by 3GPP to use unlicensed spectrum with \ac{LTE}.} However, since these proposals base on \ac{OFDM}, attending the mandatory emission mask is a significant challenge. It is fundamental that the \ac{5G} network provides an operation mode that can be used to provide broadband Internet access for those living far from city centers. The main challenges in this scenario are: i) to efficiently use the \ac{CP}, once the \ac{WRAN} channels can present multipaths that are delayed by hundreds of $\mu s$ and; ii) to properly utilize empty channels, preferably at UHF bands, without interfering with incumbents. The \ac{PHY} must provide a waveform with low \ac{OOB} emissions and high spectrum efficiency in order to address the requirement of this scenario.

\ac{5G} will face complex challenges that need to be overcome in order to make all these scenarios a reality. The adaptability of the \ac{PHY} must be taken to a new level, never seen before in a wired or wireless networks. A unique, flexible waveform that can be configured to cover all other candidates as corner cases is the ambitious solution that is pursued in this paper.

\section{Gabor transform: The theory behind \IGa{waveform design}}\label{sec:Gabor}
\begin{figure}[t]
  \centering
  \includegraphics[width=8cm]{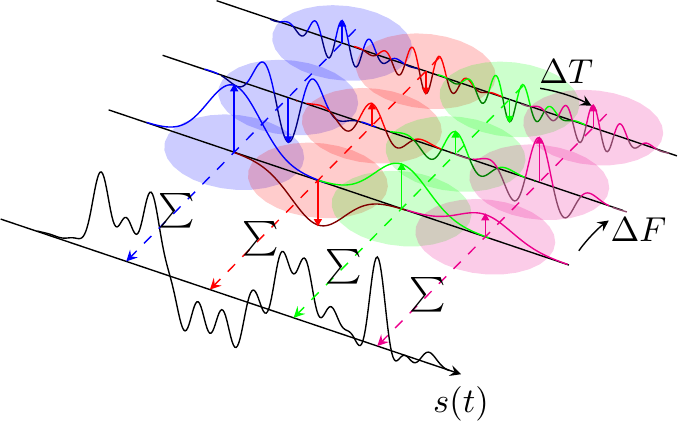}
  \caption{Illustration of Gabor expansion.  The expanded signal is the
    sum of scaled time-frequency shifts of a prototype window. The
    scaling factors are given by the Gabor expansion coefficients.}
  \label{fig:gabor}
\end{figure}

\begin{figure*}[t] 
	\centering
	\includegraphics[width=12cm]{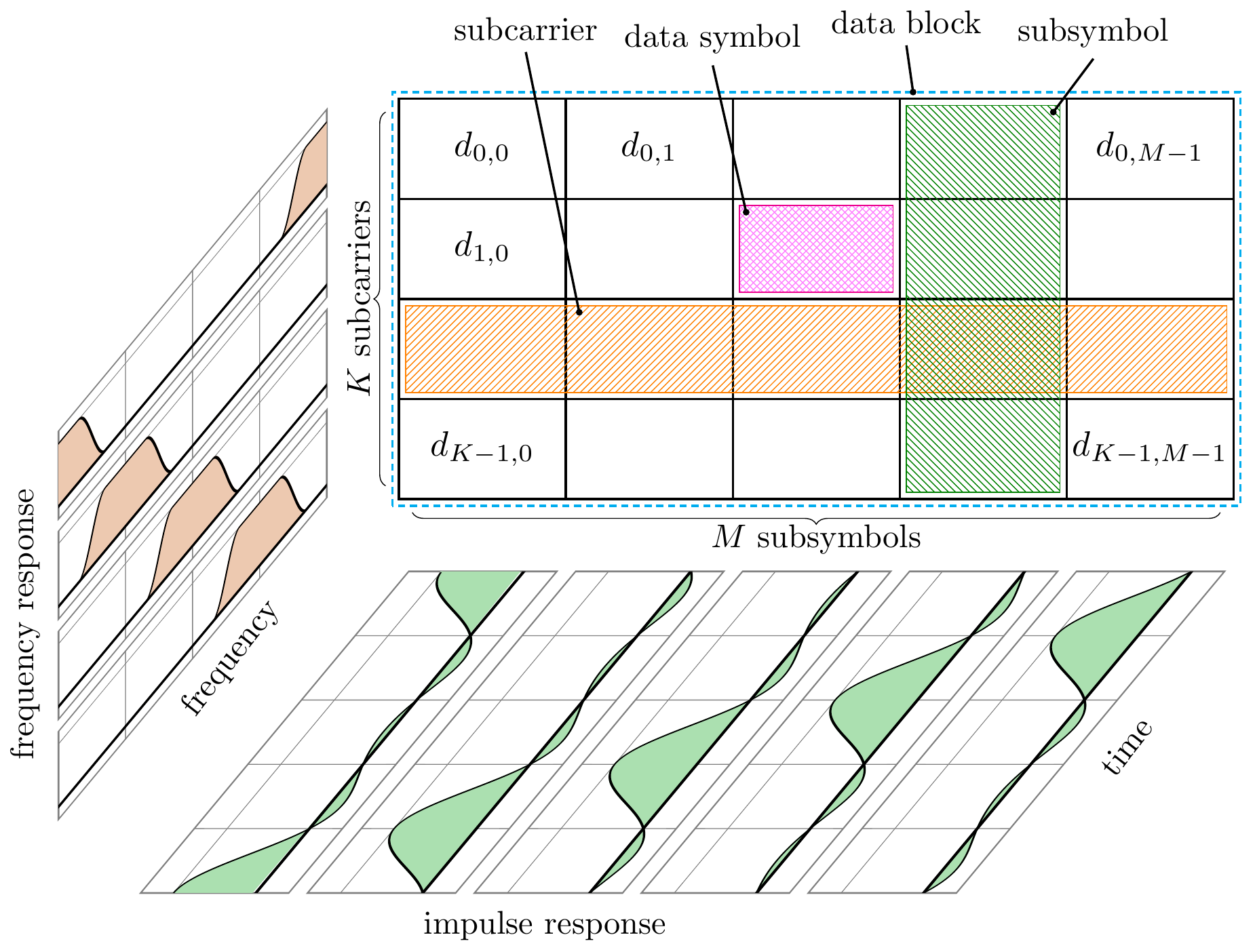}
	\includegraphics[scale=0.9]{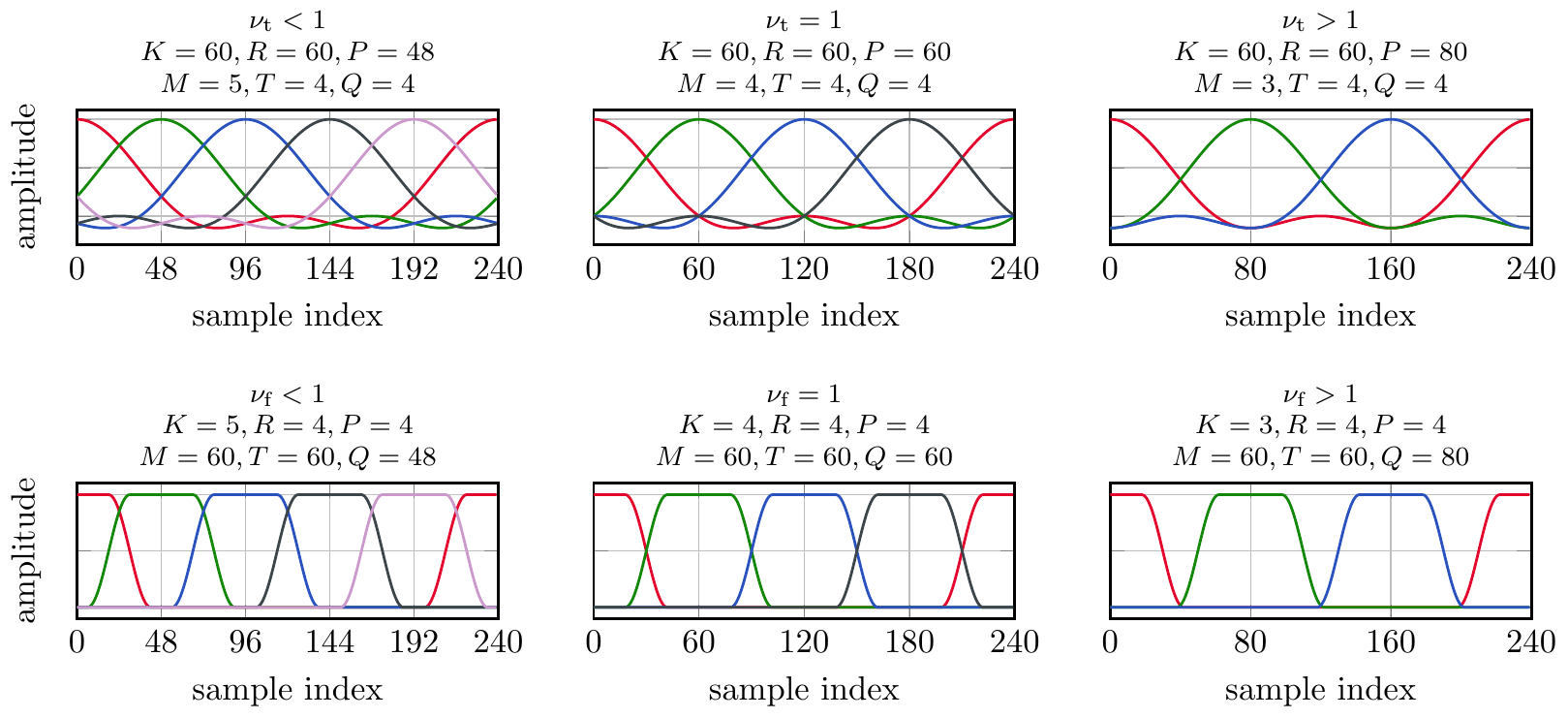}
	\caption{Illustration of the GFDM resources grid with $K=4$ subcarriers and $M=5$ subsymbols and pulse shapes, as well as subcarriers, for different values of distance factors $\nu_t$ and $\nu_f$ in time and frequency, respectively.}
	\label{fig-block}
\end{figure*} 

A software-defined \ac{BS} must rely on \ac{SDW} to achieve a flexible \ac{PHY} and the ability to explore the time and frequency dimensions is the key behind this flexibility. To better understand how these domains can be engineered, consider a signal $s$. Its time-domain representation $s(t)$ provides exact information about the behavior at any time instant. However, no information about frequency components at these positions is available. Instead, we can look at the \ac{FT} of the signal, which provides exact information about frequency components, but no information on time-domain is obtained. We can gather information about frequency components of a signal at certain time windows by looking at the FT of the multiplication of the signal with a window function, which leads to the \ac{STFT}. But the output of the \ac{STFT} can be highly redundant if the time and frequency parameters are kept independent.

To skip the unnecessary parts, in 1947, Dennis Gabor proposed to represent a signal as a linear combination of Gaussian functions that are shifted in time and frequency to positions in a regular grid, as presented in Fig. \ref{fig:gabor}.  He chose the Gaussian function because of its best localization in time and frequency simultaneously, so that local behavior of the signal is most accurately described. Gabor concluded that the original signal is fully characterized by the coefficients multiplying the Gaussian functions, establishing the foundation of time-frequency analysis \cite{Grochenig2001}. Later it was shown that the uniqueness and existence of such an expansion critically depends on the density of the grid of time-frequency shifts, which is defined as the product of spacing in time $\Delta{}T$ and frequency $\Delta{}F$. Densities larger than 1 imply non-unique expansions whereas with densities smaller than 1, expansion coefficients only exist for certain signals.

Nowadays, the linear combination of time-frequency shifted windows is known as a Gabor expansion and the calculation of the STFT with a certain window at a regular grid is known as a
Gabor transform \cite{Benedetto1998}. Expansion and transform windows are in a dual relation, i.e. the coefficients used to expand to a certain signal with a given window are provided by the Gabor transform of that signal with the dual window.  In case the window and its dual are equal, the window is said to be orthogonal and expansion and transform reduce to well-known orthogonal expansion series.

A prominent example is \ac{OFDM}, which performs a Gabor expansion using a finite discrete set of rectangular window functions with length $T_S$ in time and shifts of $1/T_S$ in the frequency grid. In discrete Gabor expansion and transform, which in the \ac{OFDM} case is the \ac{DFT}, all signals are assumed to be periodic in time and frequency.

However, non-periodic time-continuous cases can be approximated by choosing long frames and appropriate sampling frequencies. This more generic solution will be presented next as GFDM. Basically, the parameterization of the waveform directly influences i) transmitter window; ii) time-frequency grid structure; as well as iii) transform length and can hence provide means to emulate \IGa{the potential of the block based} \ac{MC} systems.

\section{\IGa{The flexible solution of a virtual \ac{PHY} layer described from the concept of GFDM}}\label{sec:GFDM}
\begin{table}[t] 
	\centering
	\caption{Terminology}
	\begin{tabular}{ll}
		\toprule
		Variable & Meaning \\
		\midrule
		$R$ & samples per period in the filter \\
		$T$ & periods in the filter \\
		$S=RT$ & total number of samples in the signal \\
		\midrule
		$P$ & subsymbol spacing in time domain \\
		$Q$ & subcarrier spacing in frequency domain \\
		\midrule
		$\nu_\text{t} = P/R$ & subsymbols distance factor \\
		$\nu_\text{f} = Q/T$ & subcarriers distance factor \\
		\midrule
		$K=RT/Q=R/\nu_\text{f}=S/Q$ & subcarriers per block \\
		$M=TR/P=T/\nu_\text{t}=S/P$ & subsymbols per block \\
		$N=KM$ & number of data symbols per block \\
		\bottomrule
	\end{tabular}
	\label{tab-params}
\end{table} 
%
%
%
Gabor theory provides the principles to use the time-frequency grid to transmit information, but it is still necessary have a modulation  scheme that can expose these resources for the upper layers in a flexible way. In all scenarios, a resource composed of a bandwidth $B_\text{W}$ and a time window $T_\text{W}$ is used to transmit a block of $N$ complex valued data symbols. With \ac{GFDM}, the physical resources are split into a two-dimensional grid, which is defined by $K$ subcarriers and $M$ subsymbols. The positions in the resource grid are denoted by the subcarrier index $k=0,\dots,K-1$ and the subsymbol index $m=0,\dots,M-1$. In each position $(k,m)$, a complex valued data symbol $d_{k,m}$ modulates the parameters of a distinct waveform $\vec{g}_{k,m}$.

We consider the case where each waveform $\vec{g}_{k,m}$ is derived from a common prototype pulse $\vec{g}$, which has a length of $S$ samples and is divided into $T$ signal periods with $R$ samples per period, i.e. $S=RT$. In principle, $\vec{g}$ can be any waveform, e.g. a raised cosine function or a rectangular function. The waveforms $\vec{g}_{k,m}$ necessary to transmit the data symbols over the resource grid are generated by circularly shifting the prototype pulse $\vec{g}$ by $mP$ samples in the time domain and $kQ$ samples in the frequency domain. For each position in the resource grid, the waveform $\vec{g}_{k,m}$ is modulated by the data symbol $d_{k,m}$. Then, all modulated waveforms are added to create a single vector that carries the information of all $N$ data symbols. The block structure of the modulated signal allows the use of a \ac{CP}, without compromising the circular signal properties of the transmitted signal.

Another aspect that can be explored with \ac{GFDM} is the density of the block structure denoted by the ratio of the number of data symbols $N$ to the number of samples in the transmitted vector $S$. In this context, it is useful to introduce $\nu_\text{t}=P/R$ and $\nu_\text{f}=Q/T$ as scaling factors for the distance between subsymbols and subcarriers, respectively. Here, values smaller than 1 lead to an increased data density.

The structure of a \ac{GFDM} block, with corresponding waveforms for the critically sampled case where $R=P$ and $T=Q$, which leads to $\nu_\text{t}=1$ and $\nu_\text{f}=1$, and different subsymbols and subcarrier distance scaling factors are illustrated in \figref{fig-block}.

More than describing a mathematical framework, in GFDM the combination of several attributes, e.g. use of guard subsymbols with pulse shaping filters to achieve both linear or circular convolution, optional guard interval with zero padding or cyclic prefix with time domain windowing, embedded unique word prefix using specific training subsymbols, frequency domain equalization to combat frequency selectivity per subcarrier, etc. can all be implemented in common software-configurable basic building blocks with hardware acceleration elements as variable length \ac{DFT}, \ac{LUT} and multiplier chains \cite{Michailow2014}.

As a 5G waveform, \IGa{circular filtered \ac{MC} systems must} fully support the application of \ac{MIMO} techniques to increase robustness and data rate. In particular, the block-based structure significantly simplifies the application of existing MIMO detection algorithms\IGa{, for instance}, in \cite{Matthe2014a} is demonstrated that spatial multiplexing can be implemented \IGa{with \ac{TR-STC}}.

\section{Offset QAM: Further increasing the flexibility}\label{sec:OQAM}

\begin{figure}[t]
	\centering
	\subfloat{\includegraphics[width=8cm]{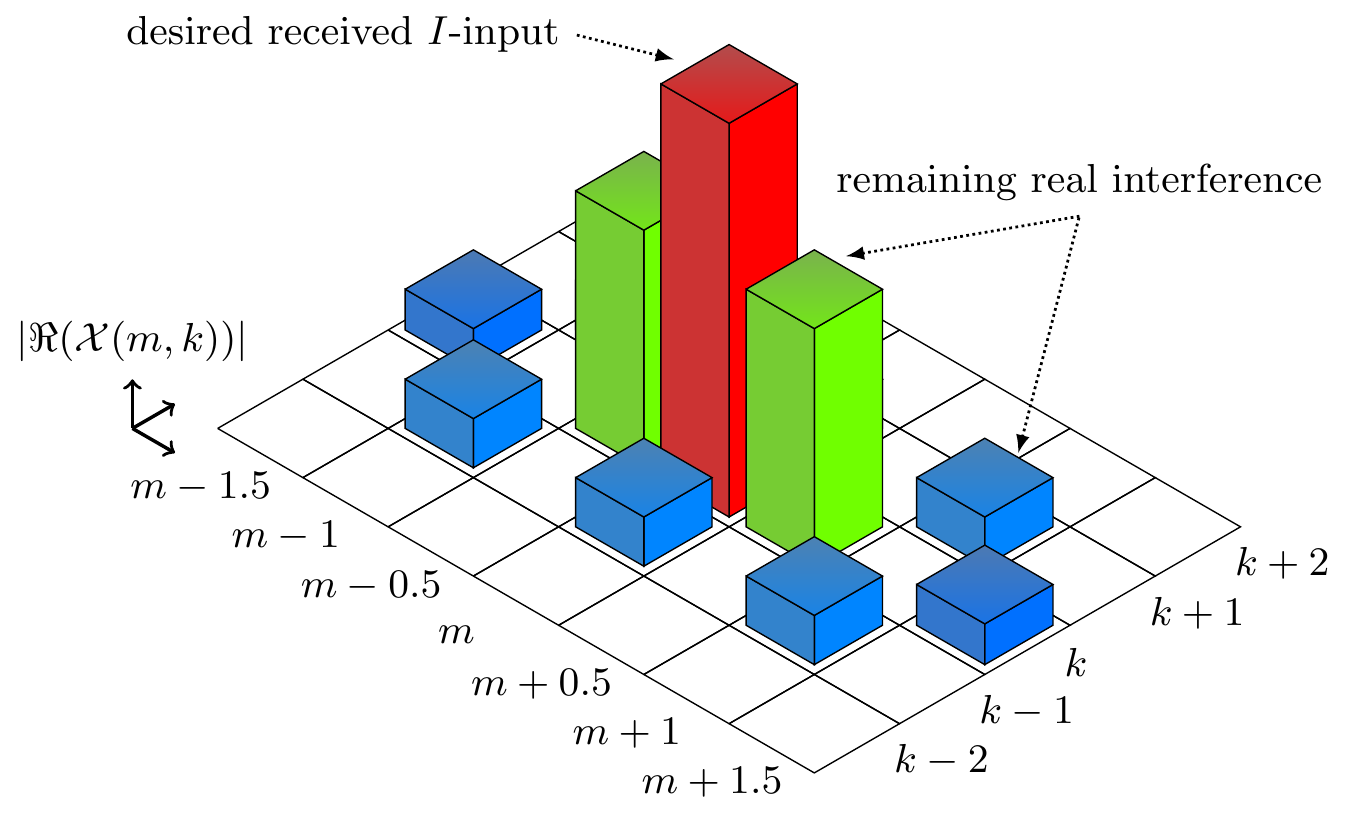}}\\
	\subfloat{\includegraphics[width=8cm]{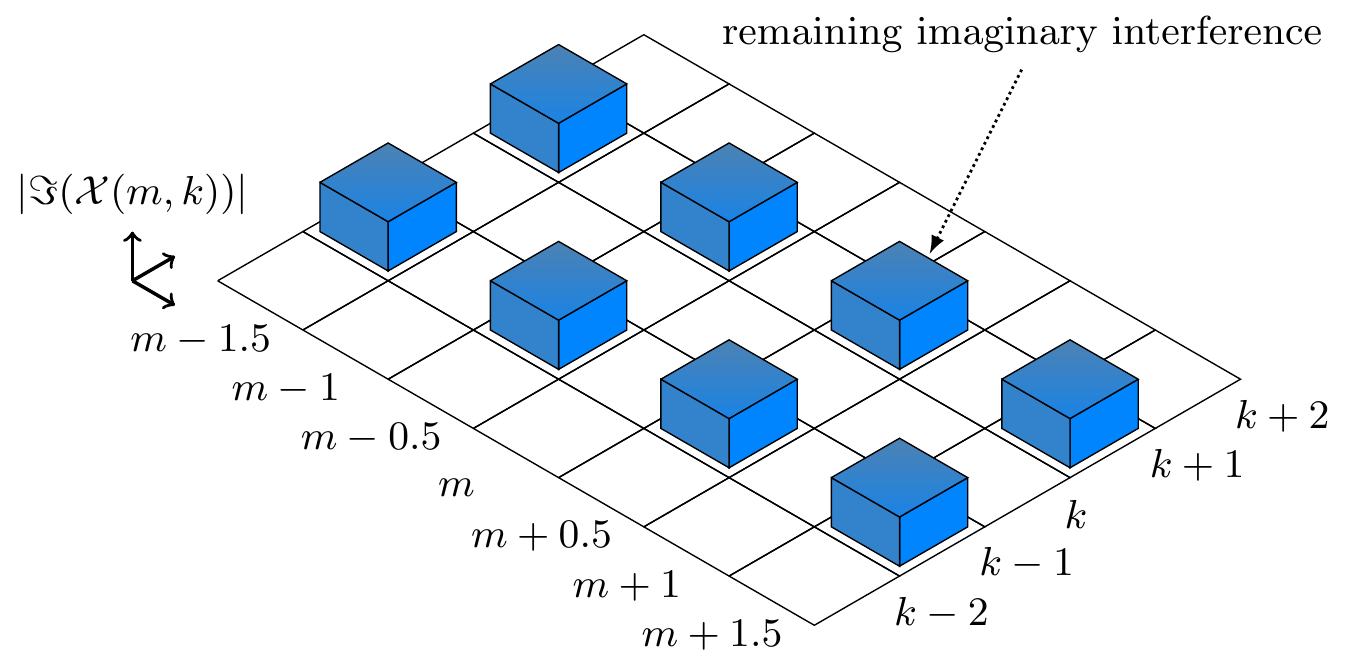}}
	\caption{Illustration of the ambiguity function $\mathcal{X}(m,k)$, with the magnitude of the residual real (top) and imaginary (bottom) values along the neighbors subsymbols and subcarriers, for a \ac{GFDM} transmitted $I$-input using RRC prototype filter with roll-off=0.5.}
	\label{fig:oqam}
\end{figure}
%
\def\sh{\shortstack} 
\def\sn{$\sqrt{\text{Nyquist}}$}
\def\nuf{$\nu_\text{f}$}
\def\nut{$\nu_\text{t}$}
\def\ms{$M_\text{s}$}
\def\mp{$M_\text{p}$}
\begin{table*} 
	\centering
	\caption{\IGa{GFDM baseline parameters and corresponding cicular filtered multicarrier associations for 5G \IGa{PHY} virtualization}}
	\begin{tabular}{l|c|*{4}{c}|*{4}{c}|*{2}{c}}
	\toprule
	\textbf{design space}	 & \textbf{GFDM} & \textbf{OFDM} & \textbf{\sh{block\\OFDM}} & \textbf{SC-FDE} & \textbf{SC-FDM} & \textbf{\sh{FBMC\\OQAM}} & \textbf{\sh{FBMC\\FMT}} & \textbf{\sh{FBMC\\COQAM}} & \textbf{CB-FMT} & \textbf{FTN} & \textbf{SEFDM} \\
	\midrule
	\# subcarriers		 & $K$		 & $K$		 & $K$		 & $1$		 & $K$		 & $K$		 & $K$		 & $K$		 & $K$		 & $K$		 & $K$		\\
	\# subsymbols		 & $M$		 & $1$		 & $M$		 & $M$		 & $M$		 & $M$		 & $M$		 & $M$		 & $M$		 & $M$		 & $1$		\\
	scaling freq.		 & \nuf		 & $1$		 & $1$		 & $1$		 & $1$		 & $1$		 & $>1$		 & $1$		 & $>1$		 & $1$		 & $<1$		\\
	scaling time		 & \nut		 & $1$		 & $1$		 & $1$		 & $1$		 & $1$		 & $1$		 & $1$		 & $1$		 & $<1$		 & $1$		\\
	silent subsym.		 & \ms		 & -		 & -		 & -		 & -		 & \mp		 & \mp		 & -		 & -		 & \mp		 & -		\\
	\midrule
	filter imp. resp.	 & cyclic	 & rect		 & rect		 & Dirichlet & Dirichlet & \sn		 & \sn		 & cyclic	 & cyclic	 & IOTA		 & rect		\\
	offset mod.			 & (yes)	 & (yes)	 & no		 & no		 & no		 & yes		 & no		 & yes		 & no		 & yes		 & no		\\
	cyclic prefix		 & yes		 & yes		 & yes		 & yes		 & yes		 & no		 & no		 & yes		 & yes		 & no		 & yes		\\
	orthogonal			 & (yes)	 & yes		 & yes		 & yes		 & yes		 & yes		 & yes		 & yes		 & yes		 & no		 & no		\\
    \midrule
   \specialcell{application \\ scenarios}   & all   & \specialcell{legacy\\systems}  & bitpipe  & IoT/MTC  & IoT/MTC  & \specialcell{WRAN,\\bitpipe} & WRAN    &  \specialcell{tactile \\ Internet}  & \specialcell{tactile \\ Internet}  & bitpipe  & bitpipe \\
   \midrule
   \specialcell{beneficial \\ features}  & flex.  & orth.   & \specialcell{small CP\\overhead}   &  \specialcell{low \\ PAPR}   & \specialcell{low \\ PAPR}   & \specialcell{low OOB}   & \specialcell{low OOB} & \specialcell{no filter \\ tail}      & \specialcell{no filter \\ tail} & \specialcell{spectral \\ eff.}    & \specialcell{spectral \\ eff.}\\
	\bottomrule
	\end{tabular}
	\label{tab-params}
\end{table*} 

Depending on the design of the pulse shaping filters, the \IGa{cyclic waveform} signal can become non-orthogonal, a situation that is not desirable in some scenarios. A flexible software-defined \ac{BS} must be able to achieve all good spectral and time properties offered by \IGa{cyclic filtered \ac{MC}}, but with the advantages provided by the orthogonality. One simple solution here is to combine \IGa{the waveform} with \ac{OQAM} \cite{gasparfrequency}.

Instead of direct modulation of the \ac{I} and \ac{Q} symbol input, the \ac{OQAM} can be seen as two independent Gabor expansions where the symbol input is restricted to be either only the real or only the imaginary component that achieves the orthogonality condition.

For square root Nyquist prototype filters with bandwidth limited to two subcarriers, the \ac{I}-input introduces only real interference in the adjacent subcarriers, while the \ac{Q}-input produces only imaginary interference. This aspect is illustrated in Fig. \ref{fig:oqam} using the ambiguity function $\mathcal{X}(m,k)$ to calculate the contribution from a transmitted symbol to every other position of the half-symbol spaced time-frequency grid. In Fig. \ref{fig:oqam}, the \ac{I}-input is used to modulate a \ac{RRC} prototype filter with roll-off $0.5$. The magnitude of real and imaginary part of $\mathcal{X}(m,k)$ reveals regions free of interference. Therefore, orthogonality can be achieved if every even and odd subcarrier is modulated with \ac{I} and \ac{Q} inputs, respectively.

Fig. \ref{fig:oqam} also shows that \ac{I}-inputs produce zero real interference in the adjacent subcarriers at every half subsymbol shift while \ac{Q}-inputs produce zero imaginary interference. Hence, a second Gabor expansion with a time-shift of half subsymbol duration can be used to transmit data free of interference.

The use of \ac{OQAM} considering pulse shaping filters with arbitrary length and overlapping factors greatly increases the flexibility of the \IGa{cyclic} waveforms. \IGa{Recently, frequency-shift \ac{OQAM} has also being introduced as an alternative for using shorter pulses in time \cite{gasparfrequency}. Also, low complexity implementation that fully explores the cyclic principles has been addressed in \cite{Michailow2014} and allows to harvest the benefits of \ac{OQAM}}.

\section{Waveform framework: GFDM as a \IGa{baseline to describe the }general \IGa{circular filtered multicarrier system} solution for 5G}\label{sec:waveforms}
%
%
%
In this section we illustrate how GFDM can be used \IGa{as a baseline to describe the overall concept of} a virtualized \ac{PHY} service, serving as a framework for multicarrier waveforms. For this purpose, the parameters and properties of \ac{GFDM} that are \IGa{needed} to \IGa{describe the complete waveform setup achieved with cyclic filtered multicarrier systems} are given in \tabref{tab-params}. All \IGa{components derived from Table \ref{tab-params}} have common roots in the filtered multicarrier systems proposed by \cite{Chang1966} 
and, being understood as special \IGa{configurations from the GFDM baseline}, are compatible its MIMO techniques \cite{Matthe2014a,Matthe2015VTC}. Although this framework covers all most known waveforms, note that some candidates are still not covered. For example, \IGa{conventional band-pass filtered \ac{CP}-\ac{OFDM} and \ac{UFMC} \cite{Wunder2014}, which applies separated linear filtering to sets of subcarriers in \ac{ZP}-\ac{OFDM} \cite{muquet2002cyclic}, are techniques that can still be combined later on with the presented block filtered schemes, replacing the \ac{OFDM} block.}

In the context of this framework, the different waveforms are characterized by two aspects. First, parameters related to the dimensions of the underlying resource grid are explored. This includes the number of subcarriers $K$ and subsymbols $M$ in the system. The scaling factor in time $\nu_\text{t}$ and frequency $\nu_\text{f}$ can theoretically take values of any rational number larger than zero, while numbers close to one are meaningful because they relate to critically sampled Gabor frames. Additionally, the option to force specific data symbols in a block to carry the value zero, i.e. so-called `guard subsymbols' \cite{Michailow2014}, with $M_\text{s}$ being a number between $0$ and $M-2$, is relevant for some candidates. The second set of features is related to the properties of the signal. Here, the choice of the pulse shaping filter is a significant attribute and the presence or absence of circularity constitutes a characteristic feature. Moreover, the use of \ac{OQAM} is needed for some waveforms, aiming to achieve higher flexibility. Further, some waveforms rely on a \ac{CP} to allow transmission of a block based frame structure in a time dispersive channel, while others don't use CP in order to achieve higher spectrum efficiency.

The family of \emph{classical waveforms} includes \ac{OFDM}, block \ac{OFDM}, \ac{SC-FDE} and \ac{SC-FDM}. Particularly \ac{OFDM} and \ac{SC-FDM} have been relevant for the development of the \ac{4G} cellular standard \ac{LTE}. All four waveforms in this category have in common that $\nu_\text{f}=1$ and $\nu_\text{t}=1$, which allows to meet the Nyquist criterion. Silent subsymbols are not employed, the \ac{CP} and regular \ac{QAM} are used used in the default configuration. \ac{OFDM} and block \ac{OFDM} are corner cases of \ac{GFDM}, where a rectangular pulse is used. Additionally, \ac{OFDM} is restricted to one subsymbol, while block \ac{OFDM} constitutes the concatenation of multiple \ac{OFDM} symbols in time to create a block with a single common \ac{CP}. Similarly, \ac{SC-FDE} and \ac{SC-FDM} can also be considered as corner cases of \ac{GFDM}. However, here a Dirichlet pulse is used and analogously, the number of subcarriers in \ac{SC-FDE} is $K=1$, while \ac{SC-FDM} is a concatenation in frequency of multiple \ac{SC-FDE} signals. All waveforms in this category share property of orthogonality, but with different sensitivities towards various \ac{RF} imperfections, for instance \ac{SC-FDE} is well known for its low \ac{PAPR}, which greatly benefits the \ac{MS} in terms of transmit power efficiency and reduced cost of the power amplifier.

The family of \emph{filter bank waveforms} revolves around filtering the subcarriers in the system and still retaining orthogonality.  As the names suggest, \ac{FBMC}-\ac{OQAM} \cite{Banelli2014} and its cyclic extension \ac{FBMC}-\acs{COQAM} \cite{Lin2014} rely on offset modulation, while in \ac{FBMC}-\acs{FMT} and \ac{CB-FMT} \cite{tonello2014cyclic} the spacing between the subcarriers is increased such that they do not overlap, i.e. $\nu_\text{f}>1$. Also, a separation between cyclic and non-cyclic prototype filters can be made. In this context, silent subsymbols become relevant. The best spectral efficiency is achieved with $M_\text{s}=0$, while  $M_\text{s}>0$ helps to improve the spectral properties of the signal. Using a sufficiently large number of silent subsymbols at the beginning and the end of a block allows to emulate non-cyclic filters from a cyclic prototype filter response, in order to generate \ac{FBMC}-\ac{OQAM} and \ac{FBMC}-\acs{FMT} bursts. More precisely, $M_\text{p}$ is the length of the prototype filter and $M_\text{s}=M_\text{p}$. Lastly, the \ac{CP} is only compatible with cyclic filters.

Generally, the waveform can become non-orthogonal depending on the use of specific filters and for a given value of $\nu_\text{f}$ and $\nu_\text{t}$. This is addressed in the final category that consists of the \emph{non-orthogonal multicarrier techniques} \ac{FTN} \cite{Banelli2014} and \ac{SEFDM} \cite{Darwazeh2009}. The key property of \ac{FTN} is $\nu_\text{t}<1$, which reflects in increment of the subsymbol data rate. The \ac{IOTA} pulse, in combination with \ac{OQAM}, has been proposed in order to avoid the need for a \ac{CP}. Since the impulse response of the filter is not cyclic, $M_\text{p}$ subsymbols are silent. Analogously, the idea of \ac{SEFDM} is to increase the density of subcarriers in the available bandwidth, i.e. $\nu_\text{f}<1$. Here, $M=1$ because each block consists of a single subsymbol that is filtered with a rectangular pulse and a \ac{CP} is prepended to combat multipath propagation. In this case, regular QAM is employed. Clearly, the amount of squeezing without severely impacting the error rate performance is limited. The Mazo limit states that this threshold is around 25\% for both schemes.

\section{Multi-service and Multi-cell \ac{5G} Networks based on \IGa{block cyclic waveforms}}
In \cite{Wunder2014}, a unified framework was proposed for delivering various services. \IGa{The waveform derivations presented so far have} good time and frequency properties to fit into \IGa{diverse 5G scenarios}. Namely, a single subcarrier as a guard band is sufficient to divide the resource grid for serving different applications in an interference free manner. Benefiting from this feature, the BS can easily deploy multiple \IGa{waveform} settings, including frame structure and pilot pattern, on a single resource grid for \ac{DL} communication. 
Notice that, although the \ac{BS} \ac{UL} receiver for a given service can be obtained through virtualization of the \ac{PHY}, the \ac{MS} just needs to have the specific transceivers for its applications. Clearly, a \ac{MS} does not need to have all possible features of GFDM implemented on its \ac{PHY}, but only the features that are requested for the services supported by this equipment.

The \IGa{cyclic waveform} time and frequency properties are also beneficial to mitigate the multi-user interference coming from asynchronous users sharing the resource grid \cite{Matthe2014a}. With a single subcarrier as the guard band, the spectrum of the different users does not overlap. As such, the time misalignment among users does not lead to multi-user interference. Since the side lobes of the subcarriers spectrum are negligible, only coarse frequency synchronization among the devices is required to avoid spectrum overlapping due to frequency misalignments. \IGa{It should be noticed that good time and frequency properties can be achieved only in combination with appropriate pulse shaping filters and, eventually needs to be combated by additional windowing, filtering, or guard/cancellation symbols.}

Inter-cell interference management will also be a challenge for 5G networks, regardless the chosen waveform and the application scenario. \ac{SDN} can enable a centralized control of \ac{BS}s and provide a global view of the network for \ac{CoMP} to align the interference and increase the overall data throughput. This approach is being pointed as the most promising solution for mitigating multi-cell interference. A \ac{CoMP} approach enormously benefits from a framework for all waveforms employed in the network.

\IGa{\section{Frame structure design: synchronization, channel estimation and advanced receivers strategies}\label{sec:rx}
The cyclic waveform can result in a simple unified transmitter, but certainly will require specific receivers architecture, since many aspects depend on the actual choice of the waveform, and the respective frame structure design. For example, channel estimation based on scattered pilots as often employed in OFDM might need to be replaced by training sequences defined in time domain to be compatible with the specific case of single-carrier transmission. Moreover, \ac{FBMC}-\ac{OQAM} or non-orthogonal waveforms might require guard intervals or iterative receivers to mitigate intrinsic interference.

To hold the waveform advantages simultaneously for all corner cases, the block based transmission can be engineered with a particularly well suited principle of pseudo circular pre/post-amble for continuous transmission over time-varying channels. An initial approach has been presented in \cite{Gaspar2015vtc}, covering synchronization aspects in vehicular communication.  

Additionally, efficient \ac{MIMO} \ac{APP} detection per subcarrier for frequency-selective channels has been initially proposed in \cite{Matthe2015VTC}, which opens the opportunity to cover most of the cyclic waveform variations presented so far in MIMO applications. In general, the scheme proposed in \cite{Matthe2015VTC} can handle not only spatial interference but also ICI/ISI, which indicates that the solutions beyond simple linear equalization or interference cancellation can be considered.
}

\section{Conclusions}\label{sec:conclusions}
%
%
\ac{NFV} is an essential paradigm in the development process of 5G networks. Our vision is to virtualize \ac{PHY} functionalities. As such, the utilization rate of network resources can be maximized by means of \ac{SDN} techniques. Additionally, the infrastructure can easily and seamlessly evolve along with the time-varying requirements. To enable virtual PHY, it is preferable and cost effective to have a single PHY relying on a software definable waveform rather than developing a miscellany of scenario dedicated \ac{PHY}s. This concept triggers the \ac{SDN} paradigms with a flexible framework for services and waveforms, which at the time of standardization eliminates the need to reserve bandwidth or resource blocks for potential future upgrades. In short, it allows not only an easy deployment and management of the network, but also provides centralized control over the resources.

\IGa{GFDM, as a baseline for circular filtered \ac{MC} systems,} have been proposed \IGa{to be an unified} \ac{SDW} to cover all other prominent candidates designed for specific scenarios. We have shown its efficient time-frequency resource grid for conveying information. This has been well exemplified in the context of Gabor transform. On the other hand, it retains necessary and sufficient degrees of freedom for waveform engineering such that the extreme requirements can be achieved by properly tuning the waveform parameters. Given this fact, we have been able to link GFDM corner cases to several waveforms in the literature that were primarily designed for individual \ac{5G} scenarios. We are aware that \ac{5G} will not be a mere evolution of \ac{4G}. Disruptive changes are expected, but legacy waveforms cannot be forgotten. As the two waveforms in the \ac{4G} \ac{PHY}, \ac{OFDM} and \ac{SC-FDM}, can be obtained from \IGa{the proposed framework} as well, legacy system can run in \ac{5G} \ac{BS} as virtual \ac{4G} \ac{BS} service. This suggests that the use of \IGa{general cyclic waveform modulation} will enable a smooth convergence of the existing \ac{4G} networks to software-defined \ac{5G} networks.

\bibliographystyle{IEEEtran}
\bibliography{library,library_Max,library_Ivan,library_LLM,library_nicola}
\end{document}